\documentclass[conference]{IEEEtran}

\usepackage[utf8]{inputenc} 
\usepackage[T1]{fontenc}
\usepackage{url}
\usepackage{ifthen}
\usepackage{cite}
\usepackage{amsmath}%
\usepackage{wasysym}%
\usepackage{amssymb}

\usepackage{mdwmath}
\usepackage{blindtext}
\usepackage{eqparbox}
\usepackage{fixltx2e}
\usepackage{stfloats}

\usepackage[english]{babel}
\usepackage{lipsum}
\usepackage{xcolor}
\usepackage{tikz}
\usepackage{mathtools,amsfonts,amssymb,amsthm}
\usepackage{graphicx}
\usepackage{algorithm, algorithmic}

\newtheorem{theorem}{{Theorem}}
\newtheorem{lemma}[theorem]{{Lemma}}
\newtheorem{proposition}[theorem]{{Proposition}}
\newtheorem{corollary}[theorem]{{Corollary}}

\newtheorem{definition}{{Definition}}

\newcommand{\cE}{{\cal E}}

\DeclareMathAlphabet{\mathbfsl}{OT1}{ppl}{b}{it} 

\newcommand{\mathset}[1]{\left\{#1\right\}}
\newcommand{\abs}[1]{\left|#1\right|}

\newcommand{\floor}[1]{\left\lfloor #1 \right\rfloor}

\def\QEDclosed{\mbox{\rule[0pt]{1.3ex}{1.3ex}}} 

\def\QED{\QEDclosed} 

\def\proof{\noindent\hspace{2em}{\itshape Proof: }}
\def\endproof{\hspace*{\fill}~\QED\par\endtrivlist\unskip}

\newcommand{\be}[1]{\begin{equation}\label{#1}}
\newcommand{\ee}{\end{equation}}

\renewcommand{\leq}{\leqslant}
 
\renewcommand{\geq}{\geqslant}

\renewcommand{\Bbb}{\mathbb}
 
\newcommand{\N}{{\Bbb N}}
\newcommand{\R}{{\Bbb R}} 
\newcommand{\Z}{{\Bbb Z}}

\newcommand{\Cref}[1]{Co\-ro\-lla\-ry\,\ref{#1}}

\newcommand{\Ftwo}{{{\Bbb F}}_{\!2}}

\newcommand{\deff}{\mbox{$\stackrel{\rm def}{=}$}}

\newcommand{\zero}{{\mathbf 0}}

\newcommand{\BSC}{{\rm BSC}}

\usepackage{stmaryrd} 
\usepackage{verbatim} 

\newcommand{\gray}[1]{\textcolor{gray}{#1}}

\usepackage{mathdots}

\IEEEoverridecommandlockouts

\begin{document}

\title{Capacity-achieving Polar-based LDGM Codes with Crowdsourcing Applications}
\author{
   \IEEEauthorblockN{\bf James (Chin-Jen) Pang, Hessam Mahdavifar, and S. Sandeep Pradhan  
   }
   \IEEEauthorblockA{ Department of Electrical Engineering and Computer Science, University of Michgan, Ann Arbor, MI 48109, USA\\
                     Email: cjpang, hessam, pradhanv@umich.edu             
                 }
    \thanks{This work was supported by the National Science Foundation
under grants CCF--1717299, CCF--1763348, and CCF--1909771.}      

 }

\maketitle

\begin{abstract}
In this paper we study codes with sparse generator matrices. More specifically, codes with a certain constraint on the weight of all the columns in the generator matrix are considered. The end result is the following. For any binary-input memoryless symmetric (BMS) channel and any $\epsilon > 2\epsilon^*$, where $\epsilon^* =  \frac{1}{6}-\frac{5}{3}\log{\frac{4}{3}} \approx 0.085$, we show an explicit sequence of capacity-achieving codes with all the column wights of the generator matrix upper bounded by $(\log N)^{1+\epsilon}$, where $N$ is the code block length. The constructions are based on polar codes. Applications to crowdsourcing are also shown. 

\end{abstract}

\section{Introduction}
Capacity-approaching error-correcting codes such as low-density parity-check (LDPC) codes \cite{gallager1962low} and polar codes \cite{arikan2009channel} have been extensively studied for applications in wireless and storage systems. Besides conventional applications of codes for error correction, a surge of new applications has also emerged in the past decade including crowdsourcing \cite{karger2011iterative,vempaty2014reliable}, distributed storage \cite{dimakis2010network}, and speeding up distributed machine learning \cite{lee2018speeding}. To this end, new motivations have arisen to study codes with sparsity constraints in their encoding and/or decoding processes. For instance, the stored data in a failed server needs to be recovered by downloading data from a few servers only, due to bandwidth constraints, imposing sparsity constraints in the decoding process in a distributed storage system. In crowdsourcing applications, e.g., when workers are asked to label items in a dataset, each worker can be assigned a few items only due to capability limitations imposing sparsity constraints in the encoding process. More specifically, low-density generator matrix (LDGM) codes become relevant for such applications \cite{mazumdar2017semisupervised,pang2019coding}. 

\subsection{LDGM and Related Works}

LDGM codes, often regarded as the dual of LDPC codes, are associated with sparse factor graphs.
The sparsity of the generator matrices of LDGM codes implies low encoding complexity. However, unlike LDPC and polar codes, LDGM code has not received significant attention. In \cite{mackay1999good, mackay1995good} it is pointed out that certain constructions of LDGM codes are not asymptotically \textit{good}, a behavior which is also studied by an error floor analysis in \cite{zhong2005approaching, garcia2003approaching}. 
Several prior works, e.g., \cite{zhong2005approaching, garcia2003approaching, zhong2005ldgm}, adopt concatenation of two LDGM codes to construct systematic capacity-approaching LDGM codes with significantly lower error floors in simulations. As a sub-class of LDPC codes, the systematic LDGM codes are advantageous for their low encoding and decoding complexity. 

In terms of the sparsity of the generator matrices, 
\cite{LDGM_capAchieving2011} showed the existence of capacity achieving codes over binary symmetric channels (BSC) using random linear coding arguments when the column weights of the generator matrix are bounded by $\epsilon N$, for any $\epsilon > 0$, where $N$ is the code block length. Also, it is conjectured in \cite{LDGM_capAchieving2011} that column weights  polynomially sublinear in $N$ suffice to achieve the capacity. For binary erasure channels (BEC), column weights being $O(\log N)$ suffice for capacity achieving, again using random linear coding arguments \cite{LDGM_capAchieving2011}. Furthermore, the scaling exponent of such random linear codes are studied in \cite{mahdavifar2017scaling}.
Later, in \cite{lin2018coding}, the existence of capacity achieving systematic LDGM ensembles over any BMS channel with the expected value of the weight of the entire generator matrix bounded by $\epsilon N^2$, for any $\epsilon > 0$, is shown.

In \cite{pang2019coding}, we formulated the problem of label learning through asking queries from  crowd workers  as a coding theory problem. Due to practical constraints in such crowdsourcing scenarios, each query can only contain a small number of items. When some workers do not respond, resembling a binary erasure channel, we showed that a combination of LDPC codes and LDGM codes gives a query scheme where the number of queries approaches information theoretic lower bound \cite{pang2019coding}.

\subsection{Our Contributions}
In this paper, we focus on studying capacity achieving LDGM codes over BMS channels with sparsity constraints on column weights. 
Leveraging polar codes, invented by Ar{\i}kan \cite{arikan2009channel}, and their extensions to large kernels, with errors exponents studied in \cite{ korada2010polar}, we show that capacity-achieving polar codes with column weights bounded by any polynomial of $N$ exist. However, a similar result can not be obtained with any polynomial of $\log N$ as the constraint on column weights. 
A new construction for LDGM codes is proposed so that most of the column weights can be bounded by a degree $1+\delta''$ polynomial of $\log N$, where $\delta''>0$ can be chosen arbitrarily small.
One issue of the new construction is the existence of, though only a few, \textit{heavy} columns in the generator matrix. In order to resolve this, we propose a splitting algorithm which, roughly speaking, splits \textit{heavy} columns into several \textit{light} columns, a process which will be clarified in the paper. The rate loss due to this modification is characterized and is shown to approach zero as $N$ grows large. Hence, the proposed modification leads to capacity achieving constructions with column wights of the generator matrix upper bounded by $(\log N)^{1+\epsilon}$, for any $\epsilon > 2\epsilon^*$, where $\epsilon^* =  \frac{1}{6}-\frac{5}{3}\log{\frac{4}{3}} \approx 0.085$. 




In crowdsourcing applications, building upon the model in \cite{pang2019coding}, we consider a scenario where some workers are not reliable, i.e., their reply to the query is not correct, each with a certain probability independent of others. We show that the LDGM codes presented in this paper in concatenation with LDPC codes can be used as query schemes where the number of queries approaches information theoretic lower bound and the number of items in each query is polylogarithmic in the number of items.

\section{Preliminaries}
\subsection{Channel Polarization and Polar Codes}\label{Prelim:polar}

The \emph{channel polarization} phenomenon was discovered by Ar{\i}kan \cite{arikan2009channel} and is based on a $2 \times 2$ polarization transform as the building block. For $N=2^n$, the polarization transform is obtained from $N \times N$ matrix $G_2^{\otimes n}$, where $G_2=\begin{bmatrix}
		1 & 0 \\
		1 & 1 \\
		\end{bmatrix}$~\cite{arikan2009channel}. 
Polar codes of length $N$ are constructed by selecting certain rows of $G_2^{\otimes n}$. More specifically, let $K$ denote the code dimension. Then sort all the $N$ bit-channels, resulting from the polarization transform, with respect to their probability of error, select the best $K$ of them with the lowest probability of error, and then select the corresponding rows from $G_2^{\otimes n}$. In other words, the generator matrix of an $(N,K)$ polar code is a $K \times N$ sub-matrix of $G_2^{\otimes n}$. The probability of error of this code, under successive cancellation decoding, is upper bounded by the sum of probabilities of error of the selected $K$ best bit-channels \cite{arikan2009channel}. Polar codes and polarization phenomenon have been successfully applied to a wide range of problems including data compression~\cite{Arikan2,abbe2011polarization}, broadcast channels~\cite{mondelli2015achieving,goela2015polar}, multiple access channels~\cite{STY,MELK}, physical layer security~\cite{MV,andersson2010nested}, and coded modulations \cite{mahdavifar2015polar}. 

\subsection{General Kernels and Error Exponent}\label{Prelim:polarExp}
It is shown in \cite{korada2010polar} that if $G_2$ is replaced by an $l \times l$ polarization kernel $G$, then polarization still occurs if and only if $G$ is an invertible matrix in $\mathbb{F}_2$ and it is not upper triangular. Furthermore, the authors of \cite{korada2010polar} provided a general formula for the error exponent of polar codes constructed based on an arbitrary $l\times l$ polarization matrix $G$. More specifically, let $N= l^n$ denote the block length and $C$ denote the capacity of the channel. For any $\beta < E(G)$, specified next, the rate $\frac{K}{N}$ of the polar code with probability of error $P_e$ upper bounded by 
$$
P_e(n) \leq 2^{-N^{\beta}}
$$
approaches $C$ as $n$ grows large. The rate of polarization (defined in \cite[Definition 7]{korada2010polar}), $E(G)$, is given by 
\begin{equation}
	E(G) = \frac{1}{l}\sum_{i=1}^l \log_l D_i, \label{eq:EGformula}
\end{equation}
where $ \mathset{D_i}_{i=1}^l$ are the partial distances of $G$. 
Formally, for $G= [g_1^T, g_2^T, \ldots, g_l^T]^T$, the partial distances $D_i$ are defined as follows:
 \begin{align}
    &D_i \deff d_H(g_i, \mbox{span}(g_{i+1}, \ldots, g_l) ), \qquad i= 1,2,\ldots, l-1 \label{eq:DiDef}\\
    &D_l \deff d_H(g_l, 0)= w_H(g_l) \label{eq:DlDef}, 
\end{align} where $d_H(a,b)$ is the Hamming distance between two vectors $a$ and $b$, and $d_H(a, U)$ is the minimum distance between a vector $a $ and a subspace $U$, i.e., $d_H(a, U)= \min_{u\in U}d_H(a,u)$.

\section{Constructions and Main Results}
The main results of this paper are stated in this section. The proofs can be found in Section \ref{sec:Appendix}.
\subsection{Sparsity Study}\label{sec:sparsityConstraint}

Leveraging results in polar coding theory, we first show the existence of capacity achieving polar codes with generator matrices of which all column weights are polynomial in the block length $N$, hence validating\gray{(right word?)} the conjecture in \cite{LDGM_capAchieving2011}. 
Second, we show that, for any polar code, \textit{almost} all of the column weights of the generator matrix are larger than polylogarithmic in $N$.
\begin{proposition}\label{prop:polarWithPolyColumnWeight}
	For  any fixed $s>0$, there are capacity-achieving polar codes with generator matrices having column weights bounded by $N^s$.
\end{proposition} 
    \begin{proposition}\label{prop:noLogColumns}
	Given $l \geq 2 $ and an $l\times l $  polarizing kernal $G$,
	the ratio of columns in $G^{\otimes n}$ with $O({(\log N)}^r)$ Hamming weight vanishes for any $r>0$ as $n$ grows large.
\end{proposition}

\subsection{New Approach: Construction}\label{sec:ConstructionNewApproach}
    We propose a new construction of codes with even sparser generator matrices than those given in section  \ref{sec:sparsityConstraint}. In particular, \textit{almost all} the column weights of the generator matrices of such codes are  logarithmic in the code block length, and there is an upper bound $w_{u.b.}$, polynomial in the logarithm of the block length, on \textit{all} the column weights. 
    
    Formally, let $G= G_l^{\otimes n}\otimes I_{n'} $, where $G_l $ is an $l\times l$ 
    polarization kernel
    and $I_{n'} $ is the $n' \times n' $ identity matrix. 
    The matrix has the following form: 
\begin{equation}
G=
  \begin{bmatrix}
    G_l^{\otimes n}       & \zero_{l^n} & \zero_{l^n} & \dots & \zero_{l^n} \\
    \zero_{l^n}       & G_l^{\otimes n} & \zero_{l^n} & \dots & \zero_{l^n} \\
    \zero_{l^n} & \zero_{l^n}      & G_l^{\otimes n}  & \dots & \zero_{l^n} \\
    \vdots & \vdots & \vdots & \ddots & \vdots \\
    \zero_{l^n}       & \zero_{l^n} & \zero_{l^n} & \dots & G_l^{\otimes n}
  \end{bmatrix}. \label{eq:Gdef}
\end{equation}
Let $N= l^n$,  $N' = N\times n'$ be the block length, and $K'= n'K$ be the code dimension.
Then $\frac{K'}{N'}= \frac{K}{N}$ is the code rate. To construct the polar-based code, we use the $K'$ bit-channels with the lowest probability of error and the generator matrix of an $(N',K')$ code \textit{based on} $G$ is a $K' \times N'$ sub-matrix of $G$.

When all columns are required to have low Hamming weights, a \textit{splitting algorithm} is applied. 
Given a column weight threshold  $w_{u.b.}$, the splitting algorithm splits any column in $G$ with weight exceeding $w_{u.b.}$ into columns that  sum to the original column both in $\Ftwo$ and in $\R$, and that have weights no larger than $w_{u.b.}$. 
That is, for a column in $G$ with weight $W$, if $W \leq w_{u.b.} $, keep the column as it is. If $W= m\cdot w_{u.b.} + r$ for some $m\in \N$ and some $0\leq r <w_{u.b.}$, replace the column with $m+1$ columns, such that each column has no more than $w_{u.b.}$ ones. 
Denote the resulting $N'\times N'(1+R) $ matrix by $G'$. 
A new code\textit{ based on}
$G'$ selects the same $K'$ rows as the code based on $G$ to form the generator matrix, whose column weights are uniformly bounded by $w_{u.b.}$.

We demonstrate the operation of the algorithm through a toy example: assume the threshold $w_{u.b.}$ is chosen to be $1$, and the first column of an $N$-column matrix $G$ is  $(1,1, 0, \ldots, 0)^T$. Then this column will be split into two new columns, $(1,0, 0, \ldots, 0)^T$ and $(0,1, 0, \ldots, 0)^T$, called $v_1'$ and $v_1''$ here. 
If all the other columns of $G$ have weights $0$ or $1$, then resulting $G'$ will be  
\begin{equation*}
G'=[v_1', v_1'', v_2, \ldots, v_N],
\end{equation*} where $v_i$ denotes the $i$th column of $G$.

 


\subsection{New Approach: Analysis of Error Probability}\label{sec:PeForNewApproach}
First, we show that, for an appropriate chice of $n'$, codes based on $G$ have vanishing probability of error as $n$ grows large.
    Let $\beta < E(G_l) $ be given,  there are polar codes with kernel $G_l$ such that the probability of error is bounded by $2^{-N^\beta}$.
    For the code based on $G$, the probability of error is bounded, through union bound,  by $n' \cdot 2^{-N^\beta}$.
    Throughout this paper, we choose
    \begin{equation}
    n' = 2^{N^{(1-\delta)E(G_l)}}\label{eq:n'def},
    \end{equation}  for an arbitrarily small constant $\delta >0.$ We then have the following lemma.
    \begin{lemma}\label{Lemma:newG_PeBound}
    	Let $G$ be as in \eqref{eq:Gdef} and $n'$ be as in \eqref{eq:n'def}. Then for any $\beta < E(G_l) $, the rate of the code based on $G$ with the probability of error upper bounded by  $2^{-N^\beta}$ approaches $C$ as $n$ grows large.
    \end{lemma}

When the splitting algorithm is applied, we show in the following proposition that the probabilities of error of the code based on $G'$ and $G$ can be bounded in the same way.  


\begin{proposition}\label{prop:codeWithG'}
	 For any $\beta < E(G_l)$, there is a decoding scheme based on successive cancellation(SC) decoding such that the probability of error of the code based on $G'$ can be bounded by $2^{-N^\beta}$ for sufficiently large $n$. 
\end{proposition}

The block length of the code based on $G$ is 
\begin{equation}\label{eq:N'def}
N' =n' N= 2^{N^{(1-\delta)E(G_l)}}  N . 
\end{equation}
We use $ log(N')$ as \textit{sparsity benchmark} in this paper, which can be bounded by
\begin{align}
	N^{E(G_l)} &\geq log(N') =N^{(1-\delta)E(G_l)}+\log{N} \nonumber \\
	&= N^{(1-\delta)E(G_l)+o(1)} \geq  N^{(1-\delta)E(G_l)}, \label{eq:logN'formula}
\end{align}{for sufficiently large $n$.}

%


\subsection{Most Common and Maximum Column Weight}
The column weights of $G$ compared to $log(N')$ can be analyzed in two scenarios: (1) most common column weight, and (2) maximum column weight. 

\begin{definition}
For a binary matrix $G$ with $m$ columns, whose weights are denoted by $w_1, w_2, \ldots, w_m $, the most common column weight $ w_{MC}(G) $ and the maximum column weight $ w_{max}(G) $ are defined as follows:
	\begin{align}
	 w_{MC}(G) &\deff \arg\max_{W}\abs{\{ i: w_i= W, 1\leq i \leq m   \}},\\
	 w_{max}(G) &\deff  \max_{i}{w_i}.
	\end{align}  
\end{definition}

Let $w_1, w_2, \ldots , w_l$ denote the column weights of the  $l\times l$ binary matrix $G_l$. The most common column weight of $G =G_l^{\otimes n}\otimes I_{n'} $
equals to that of $G_l^{\otimes n}$, which is denoted by $ w_{MC}(n, G_l)$ defined as follows:
    \begin{align}
    & w_{MC}(n, G_l) \deff w_{MC}(G_l^{\otimes n}\otimes I_{n'} ).
    \label{eq:wMCdef}
\end{align}

The maximum column weight of  $G$ is the same as that of $G_l^{\otimes n}$, which is denoted by  $w_{max}(n, G_l )$ and defined as follows: 
\begin{align}\label{eq:wmaxDef}
w_{max}(n, G_l ) \, \deff \, w_{max}(G).
\end{align}
Note that $w_{MC}(n, G_l) = [(w_1\times w_2 \times \ldots \times w_l)^\frac{1}{l}]^n = GM(w_1, w_2,\ldots, w_l)^n,$ where $GM$ is short for the geometric mean, and $n$ is assumed to be divisible by $l$. Also, $ w_{max}(n, G_l )= (max_i(w_i) )^n \leq l^n $.

\subsection{Sparsity with Kernel $G_2$}\label{subsection:mainL2}
Let $G= G_2^{\otimes n}\otimes I_{n'} $ with 
$n'$ chosen as in \eqref{eq:n'def}.
We show two things in this subsection: $w_{MC}(n, G_2) \approx \log N'$ and, after careful splitting we get a matrix $G'$ such that $w_{max}(G') \leq (\log N')^{1+2\epsilon^*}$ for a constant $\epsilon^* \approx 0.085$ with vanishing loss of rate compared to $G$. 

\begin{proposition}\label{prop:wMCloginN}
	There is a sequence of capacity achieving codes over any BMS channel with the most common column weight \textit{almost} logarithmic in the block length. More specifically, for any fixed $\delta' >0$, $n'$ in \eqref{eq:n'def} can be chosen such that 
	\begin{align}
	w_{MC}(n, G_2)=[\log(N')]^{1+\delta'+o(1)} \label{eq:wMCforG2}
	\end{align} for sufficiently large $n$. \label{prop:wMC_l=2}
\end{proposition}

By the central limit theorem, the column weights concentrate around the most common column weight, the ratio of columns with weights exceeding $[\log(N')]^{1+\delta''+o(1)}$ is vanishing as $n$ grows large for any $\delta'' >\delta'$.


Although the most common column weight of $G$ and the weights of most columns are almost logarithmic in $N'$, the maximum column weight is $w_{max}(G)= 2^n = [{w_{MC}(G) }]^2$ and is approximately ${(\log{N'})}^2$.
However, we show next that a matrix $G'$ can be obtained from the splitting algorithm such that all column weights are below some threshold $w_{u.b.}$ which would be much smaller than $w_{max}(G)$. 

Since polar codes and the code based on $G$ are capacity-achieving, as shown in lemma \ref{Lemma:newG_PeBound}, and that the code rates of the codes based on $G$ and $G'$ differ by a ratio $1+R$, the latter is capacity achieving if $R$ vanishes as $n$ grows large.
In the following, we will explore appropriate choices of the column weight threshold for the splitting algorithm that allow the value 
$R$ goes to $0$ exponentially fast. 

Let $\epsilon>0 $ be given and  \begin{equation}
    w_{u.b.}= (\log N')^{1+\epsilon}= {N}^{  \frac{1}{2}+\epsilon' },   \label{eq:wubDef}
\end{equation} be the upper bound for the column weights, where 
\begin{equation}\label{eq:epsilon'def}
    \epsilon'= (1+\epsilon ) ( \frac{1-\delta}{2}  +o(1))-\frac{1}{2},
\end{equation} for large $n$. 
 To estimate the multiplicative rate loss of $1+R$, we may study the effect on $G_2^{\otimes n}$, since that is equivalent to the overall effect on $G$. 

First note that $R$ is the ratio of the number of extra columns resulting from the splitting algorithm to the number of columns $N$ of $G_2^{\otimes n}$. Let $w_1, w_2, \ldots, w_N$ denote the column weights of $G_2^{\otimes n}$. 
 $R$ can be characterized as follows:
	\begin{equation}\label{eq:RweightCount_1}
	R=\frac{1}{N} \sum_{k=1}^{k_{max}}
	\abs{ 
\{		i:  k\cdot w_{u.b.}+1 \leq w_i < (k+1)\cdot w_{u.b.}    \}
	} \times k,
	\end{equation}where $k_{max}=\floor{\frac{2^n}{w_{u.b.}} }$.

Consider  an integer-valued discrete memoryless random process $\{W_n\}_{n\geq 0}$ 
with $W_0 = 1$, and  
$$  
W_{i+1}= \begin{cases}
\begin{aligned}
&W_i ,  &w.p. \quad \frac{1}{2}  \quad & \mbox{(the left child)}\\
&2 W_i , &w.p. \quad \frac{1}{2} \quad & \mbox{(the right child)}
\end{aligned} 
\end{cases}
$$
The random variable $W_n $ has the same distribution as the column weights of $G$ and $G_2^{\otimes n}$. 

Let $X_i$'s be i.i.d. Ber($\frac{1}{2}$) random variables defined by $
X_i \deff \log { \frac{W_{i}}{W_{i-1}}}$, and  $X(n) \deff \sum_{i=1}^n X_i \sim \text{Bin}(n, \frac12)$.  $R$ can be written as a sum of probability terms involving $X(n)$. 
\begin{lemma}\label{lemma:RinXn}
	The ratio $R$, characterized in \eqref{eq:RweightCount_1}, is equal to
	\begin{align}\label{eq:RinXn}
	R= 
    \sum_{k=1}^{k_{max}} \Pr(X(n)  > \log{(k\cdot w_{u.b.}  )} )
	\end{align} 
\end{lemma}

Assume $ \log{(w_{u.b.})}$ is an integer denoted by $ n_{lub}$ (otherwise use $ \floor{ \log{(w_{u.b.})} }$ and the analysis still holds). By grouping the $k_{max}$ terms in \eqref{eq:RinXn}, the ratio $R$ can be expressed as a sum of  $\log{k_{max}}= n-n_{lub}$ terms, 
as shown in the following lemma. 
\begin{lemma}\label{lemma:RasSum_ai}
$
	R= a_0+a_1+a_2 +\ldots +a_{n-n_{lub}-1}, 
$ where $a_i =   \Pr{( X(n) \geq 1+i+n_{lub})} \times 2^i  $.	
\end{lemma}

Let $\lambda(x,y) \deff -D(\frac{1}{2}+x+y || \frac{1}{2} )+ y$ for $x,y \geq 0$ and $x+y \leq \frac{1}{2},$ where  $D(p_1||p_2)$ is the Kullback–Leibler divergence between two distributions Ber$(p_1)$ and Ber$(p_2).$
We characterize the asymptotic behaviour of the terms in lemma \ref{lemma:RasSum_ai} in the following lemma. 
\begin{lemma}  \label{lemma:aiAsymp}
Let 
$\epsilon'$ be as in \eqref{eq:epsilon'def}. Then
	\begin{equation}
	a_i \doteq   \, 2^{n \lambda(\epsilon', \alpha_i)}, \label{eq:aiApprox}
	\end{equation} where  $\alpha_i = \frac{i+1}{n}$, 
	and $a_n \doteq b_n$ means that $\frac{1}{n}\log \frac{a_n}{b_n} \to 0$ as $n\to \infty$.
\end{lemma}
%
%


\begin{theorem}\label{Thm:ratelossAndEpsi'} 
For $G= G_2^{\otimes n}\otimes I_{n'} $, 
	where $n', N', w_{u.b.} $, and $\epsilon'$ are given by \eqref{eq:n'def}, \eqref{eq:N'def},  \eqref{eq:wubDef}, and \eqref{eq:epsilon'def}, 
	apply the splitting algorithm 
	to form a matrix $G'\in \{0,1\}^{N'\times N'(1+R)}$ such that $w_{max}(G') \leq w_{u.b.}$. 
	Then $R$ has the following asymptotic expression:
    \begin{align}\label{eq:Rasymp}
	R \doteq \begin{cases}
	2^{n(\epsilon^*- \epsilon')} \quad \to 0, &\mbox{if } \epsilon'> \epsilon^* \\
	2^{\lambda(\epsilon', \alpha_{max)} } \, \to \infty 
	, &\mbox{if } \epsilon' < \epsilon^*\end{cases},
\end{align}
	where $ \epsilon^* \deff  \hspace{1mm} \frac{1}{6}-\frac{5}{3}\log{\frac{4}{3}} \approx 0.085$ and $\alpha_{max}= \max_i \alpha_i$. 
\end{theorem} 
We can express the conditions in \eqref{eq:Rasymp} in terms of the relation between $\epsilon$ and $\epsilon^*$ leading to the following corollary. 
\begin{corollary} \label{Coro:ratelosAndEpsi}
	Let $n', N', \epsilon'$, $\epsilon^*$, $w_{u.b.}$, and $\alpha_{max}$ be as in theorem \ref{Thm:ratelossAndEpsi'}. Then
	\begin{align*}
	R \doteq \begin{cases}
	2^{n(\epsilon^*- \epsilon')} \quad \to 0, &\mbox{if } \epsilon> 2 \epsilon^* \\
	2^{\lambda(\epsilon', \alpha_{max})} \, \to \infty , &\mbox{if } \epsilon < 2 \epsilon^*\end{cases}.
	\end{align*}
\end{corollary}

The rate loss $1+R$ of the code based on $G'$ to the code based on $G$ can thus be made arbitrarily close to $1$ when the column weight upper bound $w_{u.b.}$ is properly chosen. Combining results in subsection \ref{sec:PeForNewApproach} and the corollary \ref{Coro:ratelosAndEpsi}, we have the following corollary:
\begin{corollary}\label{Coro:LDGMwithUniformBound}
    Let $\beta < E(G_2)=0.5 $ and $\epsilon >2\epsilon^* $be given. Then there exists a sequence of codes based on $G'$, generated by applying the splitting algorithm to $G= G_2^{\otimes n}\otimes I_{n'}$,
    with the following properties:
    \begin{enumerate}
        \item 
        The error probability is upper bounded by $2^{-N^{\beta}}$.
        \item
        The Hamming weight of each column of the generator matrix is upper bounded by $w_{u.b.}= (\log N')^{1+\epsilon}$.
        \item
        The rate approaches $C$ as $n$ grows large.
    \end{enumerate}
\end{corollary}

\subsection{Sparsity with General Kernels} \label{subsection:MainL>2}

In this subsection we consider $l \times l$ kernels $G_l$ with $l > 2$ and show the existence of $G_l$ with $w_{MC}(n, G_l)= O( ({\log{N'}})^{\lambda})$ for some $\lambda <1$. However, we do not bound $w_{max}(.,.)$ as in the case with the $G_2$ kernel. 
To characterize the most common column weight and the maximum column weight, the \textit{sparsity order} is defined as follows:



	\begin{definition}\label{def:lambdaMC}
	The sparsity order of the most common column weight is
	\begin{equation}
	\lambda_{MC}(n,G_l)  \, \deff \, \log_{\log(N')}{w_{MC}(n,G_l)} =\frac{\log {w_{MC}(n,G_l)}}{ \log {\log(N')}},
	\end{equation}
where $n'$ and $N'$ are defined in \eqref{eq:n'def} and \eqref{eq:N'def}, respectively. 
	\end{definition}
\begin{definition}\label{def:lambdaMax}
 the {sparsity order} of the maximum column weight
\begin{equation}
\lambda_{max}(n, G_l)  \, \deff \, 
\log_{\log(N')}{w_{max}(n,G_l)} =\frac{\log {w_{max}(n,G_l)}}{ \log {\log(N')}}.
\end{equation}
\end{definition}

For example, if ${w_{MC}(n,G_l)}$ (or ${w_{max}(n,G_l)} $) can be expressed in the Landau notations as $\Theta([\log N']^r) $, then $ \lambda_{MC}(n,G_l)$ (or $\lambda_{max}(n, G_l) $ ) goes to $r$ as $n$ grows large. 

We give Table \ref{table:orderGeneralL}\footnote{The limits of the sparsity orders when $n\to \infty$ are shown, hence $o(1)$ terms are neglected.} for
\[
G_3^* = \begin{bmatrix}
0   &   1   &   0\\
1   &   1   &   0\\
1   &   0   &   1
\end{bmatrix}, 
G_4^* = \begin{bmatrix}
1   &   0   &   0   &   0\\
0   &   1   &   0   &   1\\
0   &   0   &   1   &   1\\
1   &   1   &   1   &   1
\end{bmatrix},
\] and $G_{16}^*$ (the smallest $l$ with $E_l>0.5$; see \cite{korada2010polar} for explicit construction), which are the matrices achieving $E_3, E_4$ and $E_{16}$, the maximal error exponents for $l=3,4,16$, respectively. 

\begin{table}[t] 
\caption{$\lambda_{MC}$ and $\lambda_{max}$ for $G_2, G_3^*,G_4^* $ and $G_{16}^*$ as $n\to \infty$ }

\begin{tabular}{l | l | l | l}
 		& $E(G_l)$     & $\lambda_{MC}(n,G_l)$    & $\lambda_{max}(n, G_l)$   \\ [2mm]  \hline  
$G_2$ 	& 0.5              
		& $1+\delta'$  
		& $2(1+\delta')$   \\  [2mm]
		
$G_3^*$ 	&  $\frac{2}{3}\log_3{2}\approx 0.42$    
		&  $1+\delta'$    
		& $1.5 (1+\delta')$  \\  [2mm]

$G_4^*$ 	&   0.5                     
		&  
		$\approx 1.15(1+\delta')$                    
		& $\log{3}(1+\delta')$  \\  [2mm]

$G_{16}^*$ 
		&  $\approx 0.5183 $                   
		& $\approx  1.443(1+\delta')  $        
		& omitted \\
\end{tabular} 
\label{table:orderGeneralL}
\end{table}

However, the error exponent is not the only factor that determines the sparsity orders. For example, for $l=3$ and $l=4$, the matrices
\[
G_3' = \begin{bmatrix}
1   &   0   &   0\\
1   &   1   &   0\\
1   &   0   &   1
\end{bmatrix}, 
G_4' = \begin{bmatrix}
1   &   0   &   0   &   0\\
1   &   1   &   0   &   0\\
1   &   0   &   1   &   0\\
1   &   0   &   0   &   1
\end{bmatrix},
\] instead of $G_3^*$ and $G_4^*$, have the smallest sparsity orders of the most common column weight (found through exhaustive search), as shown in table \ref{table:BetterorderGeneralL}.
By central limit theorem, most column weights have similar orders over the logarithm of the block length. Therefore, if sparsity constraint is only required for almost all of the columns of the generator matrix, $G_3'$ and $G_4'$ are the more preferable polarization kernels over   $G_3^*$ and $G_4^*$, respectively.

\begin{table}[t] 
\caption{$\lambda_{MC}$ and $\lambda_{max}$ for   $G_3'$ and $G_4'$ as $n\to \infty$}
\begin{tabular}{l | l | l | l}
 		& $E(G_l)$     & $\lambda_{MC}(n,G_l)$    & $\lambda_{max}(n, G_l)$   \\ [2mm]  \hline  
$G_3'$ 	&  $\frac{2}{3}\log_3{2} \approx 0.42$    
		&  $
		\approx
		0.79(1+\delta') $    
		& $
		\approx
		2.38(1+\delta')$  \\  [2mm]

$G_4'$ 	&  $ \frac{3}{8}= 0.375  $        
		&  $\frac{2}{3}(1+\delta')$       
		& $\frac{8}{3}(1+\delta')$  \\
\end{tabular} 
\label{table:BetterorderGeneralL}
\end{table}


For a given $G_l$, we may relate the two terms $E(G_l)$ and $w_{MC}(n)$, or, more specifically, the partial distances $ D_1, \ldots, D_l$ and the column weights $w_1, \ldots, w_l $ as follows.

\begin{lemma}\label{lemma:SparsityOrderGeneralL}
The ratio of $\lambda_{MC}(n,G_l)$ to $\frac{ \sum_{i=1}^l \log_l w_i } {\sum_{i=1}^l \log_l D_i }$ lies between $1$ and $\frac{1}{1-\delta}$ for sufficiently large $n$.  
\end{lemma}

The following theorem shows that an arbitrarily small order can be achieved with a large $l$ and some $G_l$. 

\begin{theorem}\label{Thm:lambdaMC_lgeneral}
	For any fixed constant $0 < r\leq 1$, there exist  an $l\times l$ polarizing kernel $G_l$, where $l= l(r,\delta)$, such that $\lambda_{MC}(n,G_l) <r  $ for sufficiently large $n$.
\end{theorem}

Let $r<1$ and $\eta >0 $ be fixed. For a proper choice of  $G_l$ with $\lambda_{MC}(n,G_l) <r$, concentration of the column weights, i.e., the central limit theorem, implies only vanishing fraction of columns in $G$ have weight larger than $ [{\log{N'}}]^{(1+\eta)r}$. 

\section{Application to Crowdsourcing}

\subsection{Recap of Coding for Crowdsourced Label Learning}\label{sec:ITWrecap}
The problem model considered in \cite{pang2019coding} is the following. There are $n$ items, each of which  is associated with a binary label $X_i$ unknown to a taskmaster and $X_i$ is i.i.d. $\sim \text{Ber}(p), \forall i.$

Let $H_b(\cdot)$ denote the binary entropy function. From \cite{pang2019coding}, when workers in the crowd are perfect, there exists a XOR-querying scheme using 
$$
m= n[H_b(p)+\zeta(1-H_b(p))]
$$
queries, each involving no more than $({H_b(p)}^{-1} -1 )\frac{K_1     -K_2 ln{(\zeta)}       }{1-\zeta} $ items for some $\zeta\in (0,1)$, that achieves perfect recovery. 

In the case where queries are not responded, each with a probability $r$ independent of others, the number of queries is lower bounded by $m_{BER} = n(H_b(p))/(1-r)$ \cite{pang2019coding}. Also, existence of a XOR-querying scheme with
$$
m = n[H_b(p)+\zeta(1-H_b(p))]/(1-r)
$$
queries, each with $O( \log \frac{1}{\zeta} \log n) $ items, that guarantees perfect recovery of the labels as $n$ grows large is shown in \cite{pang2019coding}.

\subsection{BSC scenario}\label{sec:BSCmodel}
The case when some queries are answered incorrectly is widely observed in crowdsourced label learning in the real world \cite{karger2014budget, vempaty2014reliable}. When the queries are answered correctly with probability $1-q$ for some $q\in [0,0.5)$, referred to here as the $BSC(q)$ model, the information-theoretic lower bound on the number of queries is 
$$
m_{BSC}(n,p,q) = \frac{n H_b(p)}{1-H_b(q)}.
$$
We can apply corollary \ref{Coro:LDGMwithUniformBound} to design a query scheme with number of queries, $m'$, arbitrarily close to $m_{BSC}$ and small number of items in each query. 
\begin{theorem}\label{Thm:queryBSC}
	For the $BSC(q)$ model, for any $\zeta\in (0,1)$ and $\epsilon >2\epsilon^*$, there is a query scheme using 
\begin{equation}\label{eq:mforBSC}
		m' =(1+o(1))\frac{H_b(p)+\zeta(1-H_b(p))}{1-H_b(q)}
\end{equation}
	queries, each involving no more than $O(  \log \frac{1}{\zeta} {[\log n]}^{1+\epsilon})$ items,
that achieves perfect recovery. 
\end{theorem}




\section{Appendix}\label{sec:Appendix}

\subsection{Proofs for Subsection \ref{sec:sparsityConstraint}}

\textit{{Proof of Proposition \ref{prop:polarWithPolyColumnWeight}}: }

Consider an $l\times l$ polarizing matrix
	\begin{equation*}
		G=
		\begin{bmatrix}
		I_{\frac{l}{2}} & 0_{\frac{l}{2}} \\
		I_{\frac{l}{2}} & I_{\frac{l}{2}} \\
		\end{bmatrix}, 
		\label{eq:polyGform}  
	\end{equation*} where $l$ is an even integer such that $l \geq 2^{\frac{1}{s}}$.
By equations \eqref{eq:DiDef} and \eqref{eq:DlDef}, we have 
$ D_i= 1$ for $1\leq i\leq \frac{l}{2}$ and $D_i= 2$ for $\frac{l}{2}+1 \leq i\leq l$. Hence the rate of polarization 
	$E(G) = \frac{1}{2}\log_l 2  >0 $ and the polar code using $G$ as the polarizing kernel is capacity achieving. 
    For $G$, each column has weight at most $2$, so the column weights of $G^{\otimes n}$ is upper bounded  by $2^n.$
    By the choice of $l$, we know
    \[
    2^n \leq {(l^s)}^n = {(l^n)}^s = N^s. 
    \]
\endproof
	


\textit{{Proof of Proposition \ref{prop:noLogColumns}}: }

    Since $G$ is polarizing, there is at least one column in $G$ with weight at least $2$. 
    (Invertibility of $G$ implies that all rows and columns are nonzero vectors. If all column weights of $G$ are $1$, then so are all the rows, i.e., $G$ is a permutation matrix. Then $D_i =1, \forall i$, and $ E(G) =0$. Hence, $G$ is not polarizing.)
    
    Call the columns with weight larger than $1$ by\textit{ non-unity-weight} columns. Let the number of non-unity-weight columns in $G$ be $k \geq 1$.
    Let \textbf{v} be a uniformly randomly chosen column of $G^{\otimes n}$, and $w(\textbf{v}) $ be the Hamming weight of \textbf{v}. 
    For $r>0$,
    \begin{align*}
        &\Pr ( w(\textbf{v}) =  O({(\log N)}^r)  )\\
        &\leq \Pr \Big(2^{\sum_{i=1}^n F_i}   = O({(\log N)}^r)= O(  ({\log l^n} )^r  ) = O(n^r) \Big), 
    \end{align*}where $F_i $ is the indicator function that one of $k$ non-unity-weight columns in the $i$th {application} of $G$ is used to form \textbf{v}. 
    $F_1, F_2, \ldots, F_n$ are i.i.d. as $\text{Ber}(k/l) $. 
    Central limit theorem implies that
    $\sum_{i=1}^n F_i = \Theta(n)$ with high probability.
    Thus, $$ \Pr \Big(2^{\sum_{i=1}^n F_i} = O(n^r) \Big) \to 0, $$ for any $r>0$ as $n \to \infty .$
\endproof

\subsection{Proofs for Subsection \ref{sec:PeForNewApproach}}

\textit{Proof of Lemma \ref{Lemma:newG_PeBound}:}

	Let $P_e$ denote the block error probability. 
	For $\beta' < \beta < E(G_l) $, $P_e \leq 2^{-N^\beta}  $ implies $P_e \leq 2^{-N^{\beta'}}  $. Therefore, it suffices to show the probability of error bound holds for  $\beta =( 1-\eta)E(G_l)$ for any $ \eta\in (0, \delta)$. 
	
	Let $\eta$ be fixed. 
	The rate of the polar code using kernal $G_l$ with the error probability upper bounded by $2^{-N^{(1- \eta /2)E(G_l)}}$ approaches $C$ as $n$ grows large \cite{korada2010polar}. 
	For the code based on $G$, $P_e$ is bounded by 
	\begin{align*}P_e 
	&\leq n’ \times 2^{-N^{(1- \eta /2) E(G_l)}} = 2^{N^{(1-\delta)E(G_l)}} \times  2^{-N^{(1- \eta /2) E(G_l)}} \\
	&\leq 
	2^{N^{(1-\eta)E(G_l)}} \times  2^{-N^{(1- \eta /2) E(G_l)}} \\
	&=  2^{-N^{(1- \eta /2) E(G_l)} (1-N^{-\frac{\eta}{2}  E(G_l) } )  } \leq 2^{-N^{(1- \eta ) E(G_l) } } = 2^{-N^{\beta } } 
	\end{align*} {for sufficiently large} $N$.
	
	Since the rate of the code based on $G$ equals to that of the polar code using kernal $G_l$ for any $n$, it approaches $C$ as $n$ grows large.
\endproof

\textit{\\Proof of Proposition \ref{prop:codeWithG'}:}

\begin{enumerate}
\item{SC decoding: }\label{pfProp4:SCdecoding}
    
From \cite{moser2019information}, successive cancellation decoding is defined as follows:
\begin{definition}
A successive cancellation decoder decodes the bits in successive order, i.e., it firstly decodes the first information bit $\hat{U_1}$, based on the
complete channel output $Y_1^n$, then it decodes the second information bit $\hat{U_2}$ based on the complete channel output $Y_1^n$ and the already decoded first information bit  $\hat{U_1}$, etc., until at last it decodes the last information bit  $\hat{U_n}$ based on $(Y_1^n, U_1^{n-1})$.
\end{definition}

In the analysis of block error probability, however, a genie-aided successive cancellation decoding scheme, where the information of correct $U_1^{i-1}$ is available when the decoder is deciding on $\hat{U_i}$, is often more useful.
\cite[lemma 14.12]{moser2019information} states that the probabilities of error of the original SC decoder and a genie-aided successive cancellation decoder are equal. 

In terms of the generator matrix $G$ of a code, for the estimation of $\hat{U_i}$, the channel output can be thought of as the noisy version of the codeword when $U_i^{n}$ is encoded with the a matrix consisting of the bottom $n-i+1$ rows  of $G$. 

\item{Error probability bound for SC decoding: }\label{pfProp4:PeforSC}

We upper bound the probability of error of any given code scheme under successive cancellation through that of the ML decoder. 
\begin{lemma}\label{lemma:PeMLvsPeSC}
Let $P_{e,ML}$ denote the probability of block error of an $(K,N)$ code. Then the probability of error of the same code under SC decoding, denoted by $P_{e,SC}$, can be upper bounded by $K P_{e,ML}$.
\end{lemma}
\proof
Let $\cE_i$ denote the event $(\hat{U_i} \neq U_i | Y_1^N, U_1^{i-1}= \hat{U_1^{i-1}})$. 
Under the genie-aided model, \begin{align*}
P_{e,SC} = &\sum_{i=1}^K {\Pr{\cE_i}}  \leq  \sum_{i=1}^K {\Pr{ ( \hat{U_i} \neq U_i | Y_1^N ) }} \\
 &\leq \sum_{i=1}^K P_{e,ML} = K P_{e,ML} 
\end{align*}
\endproof

Since, for any $\beta < E(G_l)$, there is  a sequence of  capacity-achieving polar codes with kernel $G_l$ such that $P_{e,ML} \leq 2^{-N^\beta}$. Applying lemma \ref{lemma:PeMLvsPeSC} and similar argument as in the proof of lemma \ref{Lemma:newG_PeBound}, there exists a sequence of capacity achieving polar codes with kernel $G_l$ such that $P_{e,SC}  \leq 2^{-N^\beta}$. 

\item{Splitting on polar code improves the code:}

Let the column weight threshold $w_{u.b.}$ of $G$ be given. Let $ (G_l^{\otimes n})^{sp}$ denote the $N\times N(1+R)$ matrix generated by the splitting algorithm acting on $G_l^{\otimes n}$. 
\begin{lemma}\label{lemma:SCDforBlock}
	Under SC decoding, the probability of error of the polar code with kernel $G_l$ is no less than that of the code, with the same row indices, based on $ (G_l^{\otimes n})^{sp}$. 
\end{lemma}

\item{Decoder for the code based on $G'$:}

	When the splitting algorithm is applied on $G$, we may assume that all the new columns resulting from a column in the $j$-th chunk of $G$ are placed in the $j$-th chunk, where a chunk is the set of $N$ columns using the same $G_l^{\otimes n}$. In addition, we may require that the splitting algorithm adopts the same {division principle}. For example, the first new column includes the $w_{u.b.}$ ones with smallest row indices in the split column, the second new column includes  another $w_{u.b.}$ ones with smallest row indices, excluding those used by the first new column, and so on.
	By the structure of $G$ and the above requirement on the splitting algorithm, the matrix $G'$ has the following form: 
	\begin{equation}
	G'=
	\begin{bmatrix}
	{(G_l^{\otimes n})}^{sp}       & \zero  & \dots & \zero \\
	\zero       & {(G_l^{\otimes n})}^{sp}   & \dots & \zero \\
	\vdots & \vdots & \ddots & \vdots \\
	\zero       & \zero  & \dots & {(G_l^{\otimes n})}^{sp} 
	\end{bmatrix}, \label{eq:G'matrixForm}
	\end{equation}
	where each $\zero$ represents an $N\times N(1+R)$ zero matrix. 
	
    For the code based on $G'$, we can divide the information bits $u_1, \ldots, u_{K'}$ into $n'$ chunks, $(u_1, \ldots, u_K)$, $(u_{K+1}, \ldots, u_{2K}), \ldots, (u_{(n'-1)K+1}, \ldots ,u_{n'K}=u_{K'}   )$, written as 
    $\mathbf{u}_1, \mathbf{u}_2, \ldots, \mathbf{u}_{n'}$.
	Similarly, the coded bits $c_1, \ldots, c_{N'(1+R)}$ can be divided into $n'$ chunks, $(c_1, \ldots, c_N(1+R))$, $(c_{N(1+R)+1}, \ldots, c_{2N(1+R)})$, $\ldots, (c_{(n'-1)N(1+R)+1}, \ldots, c_{n'N(1+R)}= c_{N'(1+R)})$, denoted by 
	$\mathbf{c}_{1},\mathbf{c}_{2}, \ldots, \mathbf{c}_{n'}$. 
For the structure of $G'$ and memorylessness of the channels, the chunk $\mathbf{c}_{j}$ depends only on the information chunk $\mathbf{u}_{j}$, through a $K\times N(1+R)$ submatrix of $ {(G_l^{\otimes n})}^{sp}$, and independent of other information chunks.

We now describe the decoder for the code based on $G'$. The decoder consists of $n'$ identical copies of SC decoders, where the $j$th decoder decides on $\hat{\mathbf{u}_{j}}$ using the channel output when $\mathbf{c}_{j}$ is transmitted.

	 
\item{Error rate for the code based on $G'$:}
 
 	Lemma \ref{lemma:SCDforBlock} shows that the block-wise error probability of the code based on $ (G_l^{\otimes n})^{sp}$ is smaller or equal to that of $ G_l^{\otimes n}$. For any $ \beta < E(G_l)$, the rate of the code, whose block error probability is bounded by $2^{-N^\beta} $, approaches $C$ as $n$ grows. 
 	
    Using union bound as in lemma \ref{Lemma:newG_PeBound}, for any $ \beta < E(G_l)$, when $n'$ is chosen as in \eqref{eq:n'def}, the probability of error of the code based on $G'$ with the proposed decoder, denoted by $P_e^{SC}(G')$,  can be bounded by $P_e^{SC}(G') \leq 2^{-N^\beta} $ for sufficiently large $n$.
 	\end{enumerate}
\endproof

\textit{Proof of Lemma \ref{lemma:SCDforBlock}: }

We will show that, for an  $N_1\times N_2$ matrix $M$, when a column is split into two nonzero columns, which replaces the split column and forms an  $N_1\times (N_2+1)$ matrix $M'$, the probability of error of the code with generator matrix $M$ is lower bounded by that with $M'$.

Instead of $M$ and $M'$, we consider another two matrices below that define the same code, respectively, when used as generator matrices. 
WLOG, we may assume the first column of $M$, denoted by $v_1$,  is split into two columns and become, say, the first two columns of $M'$, denoted by $v_1', v_1'' $. We may also assume the first two elements of $v_1$ are both $1$. 
The columns are associated by the following equation:\
\begin{equation}\label{eq:v1v1'}
v_1= 
\begin{pmatrix} 1 \\ 1 \\ * \\ \vdots \\ * \\ *  \\ \vdots \\ * \end{pmatrix} = 	
\begin{pmatrix} 1 \\ 0 \\ * \\ \vdots \\ * \\ 0  \\ \vdots \\ 0 \end{pmatrix} +
\begin{pmatrix} 0 \\ 1 \\ 0 \\ \vdots \\ 0 \\ *  \\ \vdots \\ * \end{pmatrix} = 	
v_1'+ v_1''.	 		
\end{equation}			
For $v_1'$, row operations can be applied to cancel out the nonzero entries in the $*$ positions using the $1$ at the first row. Similarly, the second row of $v_1''$ can be used in row operations to cancel out any nonzero entries in the $*$ positions. Let $E$ be the matrix of the concatenation of the row operations on $v_1'$ and $v_1''$. We have 
\begin{equation}\label{eq:v1v1'}
E v_1= 
\begin{pmatrix} 1 \\ 1 \\ 0 \\ \vdots \\ 0 \\ 0  \\ \vdots \\ 0 \end{pmatrix} = 	
\begin{pmatrix} 1 \\ 0 \\ 0 \\ \vdots \\ 0 \\ 0  \\ \vdots \\ 0 \end{pmatrix} +
\begin{pmatrix} 0 \\ 1 \\ 0 \\ \vdots \\ 0 \\ 0  \\ \vdots \\ 0 \end{pmatrix} = 	
E(v_1'+ v_1''). 		
\end{equation}

Since $E$ is composed of a sequence of invertible $N_1 \times N_1$ matrices, the span of the row vectors of $M$ is the same as that of $E M$. 
Similarly, the row vectors of $M'$ and $E M'$ have  the same span space. Hence the codes with generator matrices being either $M$ or $E M$ are the same, and so are the codes with either $M'$ and $E M'$. 
We then show that the code with generator matrix $E M'$, denoted by $C_{EM'} $, is no worse than the code with $EM$, denoted by $C_{EM} $, in terms of  error probability  under SC decoding. 

Reason: Let  $u_1, \ldots, u_{N_1}$ be the information bits, $y_1,y_2$,$ \ldots$,$ y_{N_2}$ be the channel output of $C_{EM} $ and $y_1',y_1'', z_2 \ldots, z_{N_2}$ be the channel output of $C_{EM'}$. 
Note that $(y_2, \ldots, y_{N_2})$ and  $(z_2 \ldots, z_{N_2})$ are identically distributed given the information bits. Assume SC decoding outputs $\hat{u_i}$'s in increasing order of the index $i$.

As mentioned in part \ref{pfProp4:SCdecoding}, to decide on $\hat{u_i}$, we may assume that the $(N_1-i+1)\times N_2$ submatrices of $EM$ and $EM'$ are used as generator matrices for information bits $u_i, \ldots, u_{N_1}$ and the codewords are transmitted through the channel. 
For $i=2$, the submatrices of $EM$ and $EM'$ are the same for the $N_2-1$ columns from the right. The first column of the submatrix of $EM'$ is a zero vector, and the first column of the submatrix of $EM$ is equal to the second column of that of $EM'$. 
For $i>2$, the submatrices of $EM$ and $EM'$ are the same for the $N_2-1$ columns from the right, and the columns corresponding to $Ev_1, Ev_1'$, and $Ev_1''$ are zero vectors.  
Therefore, the  bit channel seen by $U_i$ is the same for both codes $C_{EM} $ and $C_{EM'}$ for $i\geq 2$.

Hence it suffices to show that probability of error in the estimate of $u_1$  with $C_{EM'} $  is not worse than that with $C_{EM}.$

Consider two channels: 
$W$ with input $U_1$ and output $Y_1, \ldots, Y_{N_2}$ and 
$W'$ with input $U_1$ and output $Y_1',Y_1''$,$Z_1$,$\ldots$, $Z_{N_2}$. Let $\mathbf{y}$ and $\mathbf{y}_{-1}$ denote $\mathbf{y}_1^{N_2}$ and $\mathbf{y}_2^{N_2}$, respectively. 
The probability of error of the first channel 
\begin{align}
P_e(W)= 
\frac{1}{2}
\sum_{\mathbf{y}}
\min( &W(\mathbf{y}|U_1= 1), W(\mathbf{y}|U_1= 0)  )  \nonumber \\
=\frac{1}{2}\sum_{\mathbf{y}_{-1}}
[
\min(&W(y_1= 0,y_2^{N_2}=\mathbf{y}_{-1}|0),\nonumber\\
&W(y_1= 0,y_2^{N_2}=\mathbf{y}_{-1}|1)  ) \nonumber \\
+\min(&W(y_1= 1,y_2^{N_2}=\mathbf{y}_{-1}|0),\nonumber\\
&W(y_1= 1,y_2^{N_2}=\mathbf{y}_{-1}|1)  ) ]\nonumber \\
=\frac{1}{2}\sum_{\mathbf{y_{-1}}}
[
\min(&A_{0,0}(\mathbf{y}_{-1}), A_{0,1}(\mathbf{y}_{-1})  ) \nonumber \\
+\min(&A_{1,0}(\mathbf{y}_{-1}),A_{1,1}(\mathbf{y}_{-1}) ) ],
\label{eq:PeW_eq1}
\end{align} 
where \[
A_{i,j}(\mathbf{y}_{-1}) \deff W(y_1= i,y_2^{N_2}=\mathbf{y}_{-1}|U_1= j).
\]

Defining
\begin{align*}
    B_{0,0}(\mathbf{y}_{-1})&=\Pr{(Y_2^{N_2}=\mathbf{y}_{-1} | U_1= 0, U_2= 0)},\\
    B_{0,1}(\mathbf{y}_{-1})&=\Pr{(Y_2^{N_2}=\mathbf{y}_{-1} | U_1= 0, U_2= 1)},\\
    B_{1,0}(\mathbf{y}_{-1})&=\Pr{(Y_2^{N_2}=\mathbf{y}_{-1} | U_1= 1, U_2= 0)}, \\
    B_{1,1}(\mathbf{y}_{-1})&=\Pr{(Y_2^{N_2}=\mathbf{y}_{-1} | U_1= 1, U_2= 1)},
\end{align*} the terms in \eqref{eq:PeW_eq1} can be written as ($y_2^{N_2}$ is omitted for simplicity)
\begin{align*}
    2A_{0,0}&=pB_{0,1} +(1-p)B_{0,0} , \\
    2A_{0,1}&=pB_{1,0} +(1-p)B_{1,1} , \\
    2A_{1,0}&=pB_{0,0} +(1-p)B_{0,1} , \\
    2A_{1,1}&=pB_{1,1} +(1-p)B_{1,0}. 
\end{align*}

We can also write down $P_e(W')$ as 
\begin{align}
    P_e(W')
    =\frac{1}{2}\sum_{\mathbf{y}_{-1}}[
    &\min{(W'(0,0,y_2^{N_2}|0),W'(0,0,y_2^{N_2}|1)  )}\nonumber\\
    +&\min{(W'(0,1,y_2^{N_2}|0),W'(0,1,y_2^{N_2}|1)  )}\nonumber\\
    +&\min{(W'(1,0,y_2^{N_2}|0),W'(1,0,y_2^{N_2}|1)  )}\nonumber\\
    +&\min{(W'(1,1,y_2^{N_2}|0),W'(1,1,y_2^{N_2}|1)  )}
    ]\nonumber\\
    = \frac{1}{2}\sum_{\mathbf{y}_{-1}}
    [
    &\min{((1-p)A_{0,0}(\mathbf{y}_{-1}),pA_{1,1}(\mathbf{y}_{-1}))}  \nonumber\\
    +&\min{((1-p)A_{1,0}(\mathbf{y}_{-1}),pA_{0,1}(\mathbf{y}_{-1}))}     \nonumber\\
    +&\min{((1-p)A_{1,1}(\mathbf{y}_{-1}),pA_{0,0}(\mathbf{y}_{-1}))} \nonumber\\
    +&\min{((1-p)A_{0,1}(\mathbf{y}_{-1}),pA_{1,0}(\mathbf{y}_{-1}))} 
    ]\label{eq:PeW'terms}
\end{align}

We show $P_e(W') \leq P_e(W)$ by showing that for each $\mathbf{y}_{-1}$, the summand in \eqref{eq:PeW'terms} are smaller or equal to that in \eqref{eq:PeW_eq1}.

We consider 3 cases for different orders of the $A_{i,j}$ terms.
\textit{Case 1:}

Assume $ A_{0,1},A_{1,0}$ are neither the largest two nor the smallest two among $A_{0,0}, A_{0,1},A_{1,0}, A_{1,1}.$ (For example, $ A_{0,1}> A_{0,0} > A_{1,1} >A_{1,0}$ or $ A_{0,1}> A_{0,0}  >A_{1,0} > A_{1,1} $  are possible orders.)

Recall that for any $a,b,c,d\in \R$, the inequality \[
\min{(a,b)} +\min{(c,d)} \leq  \min{(a+c, b+d)}
\] always holds. 
Terms in \eqref{eq:PeW'terms} can be upper bounded as follows:
\begin{align*}
&\min{((1-p)A_{0,0},pA_{1,1})}  
+\min{((1-p)A_{1,1},pA_{0,0})} \nonumber\\
\leq &\min{( A_{0,0},A_{1,1})},\mbox{ and} \nonumber \\
&\min{((1-p)A_{1,0},pA_{0,1})}  
+\min{((1-p)A_{0,1},pA_{1,0})} \nonumber\\
\leq &\min{( A_{0,1},A_{1,0})}. \nonumber
\end{align*}
The assumption guarantees the minimum of $\{A_{0,0},A_{1,1}\} $ and the minimum of $\{A_{0,1},A_{1,0}\} $ are the smallest two among $\{A_{0,0}, A_{0,1},A_{1,0}, A_{1,1}\}.$ Therefore the summand of \eqref{eq:PeW'terms} is no larger than the sum $\min{( A_{0,0},A_{1,1})}+ \min{( A_{0,1},A_{1,0})}$, which is a lower bound of the summand in \eqref{eq:PeW_eq1}. 

\noindent\textit{Case 2:}

Assume $\{A_{0,1},A_{1,0}\} $ are larger than $\{A_{0,0},A_{1,1}\} $. 
Then $A_{1,0} \geq  A_{1,1}$ and $A_{0,1} \geq  A_{0,0}$, thus 
\begin{align*}
&pB_{0,0} +(1-p)B_{0,1} \geq pB_{1,1} +(1-p)B_{1,0} \\
&pB_{1,0} +(1-p)B_{1,1} \geq pB_{0,1} +(1-p)B_{0,0}.
\end{align*}
Using $p\leq \frac{1}{2}$, 
\begin{align}
&(1-p)B_{0,0} +(1-p)B_{0,1} \geq pB_{0,0} +(1-p)B_{0,1} \nonumber \\ 
&\geq pB_{1,1} +(1-p)B_{1,0} \geq pB_{1,1} +pB_{1,0}, \mbox{and } \label{pfLemmaWW':Bineq1} \\
&(1-p)B_{1,0} +(1-p)B_{1,1} \geq pB_{1,0} +(1-p)B_{1,1} \nonumber \\
&\geq pB_{0,1} +(1-p)B_{0,0} \geq pB_{0,1} +pB_{0,0}. \label{pfLemmaWW':Bineq2}
\end{align}

The terms in \eqref{eq:PeW'terms} can be upper bounded as follows:
\begin{align*}
&\min{((1-p)A_{0,0},pA_{1,1})}  
+\min{((1-p)A_{1,0},pA_{0,1})} \nonumber\\
\leq &\min{( (1-p)( A_{0,0}+A_{1,0}),p(A_{1,1}+ A_{0,1}))}\\
= \frac{1}{2}&\min{( (1-p)( B_{0,1}+B_{0,0}),p(B_{1,1}+ B_{1,0}))}\\
= \frac{1}{2}&p(B_{1,1}+ B_{1,0})
\mbox{ by \eqref{pfLemmaWW':Bineq1}, and} \nonumber \\
&\min{((1-p)A_{1,1},pA_{0,0})}  
+\min{((1-p)A_{0,1},pA_{1,0})} \nonumber\\
\leq &\min{( (1-p)( A_{1,1}+A_{0,1}),p(A_{0,0}+ A_{1,0}))}\\
= \frac{1}{2}&\min{( (1-p)( B_{1,1}+B_{1,0}),p(B_{0,1}+ B_{0,0}))}\\
= \frac{1}{2}&p(B_{0,1}+ B_{0,0}) \mbox{ by \eqref{pfLemmaWW':Bineq2}}.
\end{align*}

The summand in \eqref{eq:PeW_eq1} can be written as:
\begin{align*}
&\min{( A_{0,0}, A_{0,1})} +\min{(A_{1,0},A_{1,1} )}  \\
= & A_{0,0}+ A_{1,1} \\
= &\frac{1}{2}[ pB_{0,1} +(1-p)B_{0,0} +pB_{1,1} +(1-p)B_{1,0}  ] \\
\geq &\frac{1}{2}p[B_{0,1} +B_{0,0} +B_{1,1} +B_{1,0}],
\end{align*} which is larger than or equal to that of \eqref{eq:PeW'terms}.

\noindent\textit{Case 3:}

Assume $\{A_{0,1},A_{1,0}\} $ are smaller than $\{A_{0,0},A_{1,1}\} $. 
Similar argument as in case 2 can be used to show that the summand in \eqref{eq:PeW_eq1} are lower bounded by that in \eqref{eq:PeW'terms}.

Therefore, the probability of error of the code with generator matrix $M$ is lower bounded by that with $M'$. By induction, the error probability of the code based on $ (G_l^{\otimes n})^{sp}$ is no larger than that of the code based on $ G_l^{\otimes n}$.  \endproof

\subsection{Proofs for Subsection \ref{subsection:mainL2}}

\textit{Proof of Proposition \ref{prop:wMCloginN}: }

The most common column weight of $G$ equals to that of  $G_2^{\otimes n}$, which is $2^{n/2}= \sqrt{N}$. The column weight benchmark is $\log(N') =  N^{(1-\delta)/2} +\log N= (N^\frac{1}{2})^{(1-\delta) +o(1)} $ for large $n$.  
Thus, for large $n$,
\begin{align*}
w_{MC}(n, G_2) &=  \sqrt{N} \hspace{-1mm} = \hspace{-1mm}[\log(N')]^{\frac{1}{1-\delta+o(1)}} \hspace{-1mm} =\hspace{-1mm}[\log(N')]^{1+\delta'+o(1)} \hspace{-3mm}.
\end{align*}
Note that $\delta>0 $ can be chosen arbitrarily small, which completes the proof.  
\endproof

\vspace{2mm}
\textit{Proof of Lemma \ref{lemma:RinXn}:}

\begin{align*}    
R    =&\Pr ( 2 w_{u.b.} \geq W >w_{u.b.} )\times 1 \nonumber \\
    &+ \Pr ( 3 w_{u.b.} \geq W >2 \cdot w_{u.b.} )\times 2\nonumber \\    
    &+\ldots +\Pr( 2^n \geq W > k_{max} \cdot w_{u.b.})\times k_{max} \nonumber\\
    =& \Pr (W >w_{u.b.}) +\Pr (W >2 w_{u.b.}) \nonumber\\
    &+ \ldots +\Pr (W > k_{max}\cdot w_{u.b.}). \nonumber
\end{align*}
Since $ 2^{X_i}= { \frac{W_{i}}{W_{i-1}}} $, we have 
$W= W_n = \frac{W_{n}}{W_{n-1}}\times \frac{W_{n-1}}{W_{n-2}}\times \ldots \times \frac{W_{1}}{W_{0}} \times W_0= 
2^{\sum_{i=1}^n X_i} = 2^{X(n)}$. 

\begin{align*}
R =& \Pr(2^{X(n)}  > w_{u.b.} )+ \ldots 
+\Pr (2^{X(n)} > k_{max}\cdot w_{u.b.}) \nonumber \\
= &\Pr(X(n)  > \log{(w_{u.b.}  )} )+ \ldots \\
&+\Pr (X(n) > \log{\Big( k_{max}\cdot w_{u.b.} \Big)}) \\
=&\sum_{k=1}^{k_{max}} \Pr(X(n)  > \log{(k\cdot w_{u.b.}  )} ).
\end{align*}
\endproof


\textit{Proof of Lemma \ref{lemma:RasSum_ai}:}

Recall that $X(n) \deff \sum_{i=1}^n X_i$ is an integer-valued random variable, so we group the terms in \eqref{eq:RinXn}  as 
\begin{align}
R	
     =&\Pr{( X(n) >n_{lub})} +	\Pr{( X(n) >\log{2}+n_{lub})}\nonumber\\
     &+\Pr{( X(n) >\log{3}+n_{lub})} +\Pr{( X(n) >\log{4}+n_{lub})}\nonumber \\
 	&+\ldots +\Pr{( X(n) >n-  n_{lub}+ n_{lub}) } \label{eq:R_lemmaPf_1}\\
= &\Pr{( X(n) >n_{lub})} \times 2^0 +\Pr{( X(n) >1+n_{lub})}\times 2^1\nonumber\\
&+ \Pr{( X(n) >2+n_{lub})}\times 2^2 +\ldots \nonumber \\
&+\Pr{( X(n) >n-1)}\times 2^{(n-n_{lub}-1)} \label{eq:R_lemmaPf_2}\\
 = & \Pr{( X(n) \geq 1+n_{lub})} \times 2^0 +\Pr{( X(n) \geq 2+n_{lub})}\times 2^1\nonumber\\
 &+ \Pr{( X(n) \geq 3+n_{lub})}\times 2^2 +\ldots \nonumber \\
 &+\Pr{( X(n) \geq (n-n_{lub})+n_{lub})}\times 2^{(n-n_{lub}-1)}\nonumber\\
= & a_0+a_1+a_2 +\ldots +a_{n-n_{lub}-1}\nonumber,   \label{eq:weightCount_3}
\end{align} where, to derive \eqref{eq:R_lemmaPf_2}, we used the fact that $X(n) > a+b$ for some  $a\in \Z$ and $b\in (0,1]$ if and only if  $X(n) >a$. 
\endproof

\vspace{2mm}
\textit{Proof of Lemma \ref{lemma:aiAsymp}:}

	Using Sanov's Theorem (\cite[Thm 11.4.1]{cover2012elements}), the $a_i $ term  in lemma \ref{lemma:RasSum_ai} can be bounded as follows\footnote{Remark: In fact, even the polynomial term in the upper bound can be dropped since the set of distribution $E$, as defined in \cite{cover2012elements}, is a convex set of distributions.}: 
	 \begin{equation*}
	\frac{1}{(n+1)^2} 2^{-n D(P_i^*  || Q  )}\cdot 2^i \leq a_i \leq (n+1)^2  2^{-n D(  P_i^*  || Q   )}\cdot 2^i, 
	\end{equation*} where $P_i^*$ and $Q$ are the $Ber( \frac{i+1+n_{lub}}{n})$ and $Ber(\frac{1}{2})$ distributions, respectively.
	For $0\leq i \leq n-n_{lub}-1 $, we have the equation
	\[a_i \doteq 2^{-n D(P_i^*  || Q  )+i+1} = 2^{n (- D(P_i^*  || Q  ) +\alpha_i)  }.\]


	
	Using \eqref{eq:wubDef}, 
	\[
	n_{lub}=
	\log{(w_{u.b.})}= \log{{N}^{ \frac{1}{2}+\epsilon' }} 
	= (\frac{1}{2}+\epsilon')n. \]
	
	Hence 
	$P_i^*$ can be written as the $Ber(\frac{1}{2}+\epsilon'+\alpha_i)$ distribution. 
\endproof

\textit{Proof of Theorem \ref{Thm:ratelossAndEpsi'}:}

The ratio $R$ is bounded by \begin{align}
\max_{i}a_i \leq R \leq n\cdot \max_{i}a_i.
\label{eq:Rbounds}
\end{align}
Therefore, the rate loss of the code based on $G'$ instead of $G$ is either vanishing, when $\max_i \lambda(\epsilon', \alpha_i)<0, $ or infinite, when there's one $i$ such that $ \lambda(\epsilon', \alpha_i)>0$.



 	The exponent can be analyzed as follows:
 	\begin{align*}
 	&\lambda(\epsilon', \alpha_i)= \alpha_i-(\frac{1}{2}+\epsilon'+\alpha_i )\cdot \log{(2 (\frac{1}{2}+\epsilon'+\alpha_i ) )}\\
 	&- (\frac{1}{2}-\epsilon'-\alpha_i )\cdot 
 	\log{(2 (\frac{1}{2}-\epsilon'-\alpha_i ) ) }\\
    &= \alpha_i- \frac{1}{2}\log{\Big(  (1+2\epsilon'+2\alpha_i)(1-2\epsilon'-2\alpha_i) \Big)}\\
    &-(\epsilon'+\alpha_i)\log{\frac{1+2\epsilon'+\alpha_i}{1-2\epsilon'-2\alpha_i} }.
 	\end{align*}
 	Consider $\lambda(\epsilon', \alpha)$ as a function of $\alpha$, we find its first and second derivatives with respect to $\alpha$:
 	\begin{align}
 	&\frac{\partial \lambda(\epsilon', \alpha)}{\partial \alpha} = 1-\log{\frac{1+2\epsilon'+2\alpha}{1-2\epsilon'-2\alpha} }, \label{eq:lambdaDeri}
 	\end{align}
 	and 
 	\begin{align}
 	&\frac{\partial^2 \lambda(\epsilon', \alpha)}{\partial^2 \alpha}\nonumber = -\frac{\partial \log ( 1+2\epsilon'+2\alpha)}{\partial \alpha} +\frac{\partial \log( 1-2\epsilon'-2\alpha)}{\partial \alpha}\nonumber\\\vspace{3mm}
 	&=-\frac{1}{\ln{2}} \frac{2}{ 1+2\epsilon'+2\alpha}+ \frac{1}{\ln{2}} \frac{-2}{ 1-2\epsilon'-2\alpha} \nonumber \\ \vspace{3mm}
 	&= -\frac{2}{\ln{2}}\big( \frac{1}{ 1+2\epsilon'+2\alpha}+ \frac{1}{ 1-2\epsilon'-2\alpha} \big) <0
 	\end{align} for any $\frac{1}{n} \leq \alpha \leq \frac{1}{2}-\epsilon'. $

 	Thus, for any fixed $\epsilon'$, $\lambda(\epsilon', \alpha)$ is a concave function of $\alpha$ and has maximum when \[\frac{\partial \lambda(\epsilon', \alpha)}{\partial \alpha} =0 . \]
 	From \eqref{eq:lambdaDeri}, the above equality holds iff $\alpha= \frac{1}{6}-\epsilon'$, and the maximum is 
 	\begin{equation}
 	\max_{\alpha}\lambda(\epsilon', \alpha) =
 	\lambda(\epsilon', \frac{1}{6}-\epsilon')= \frac{1}{6}-\frac{5}{3}\log{\frac{4}{3}} -\epsilon' = \epsilon^*- \epsilon'
 	\label{maxLambda}
 	\end{equation}

 	If $\epsilon'> \epsilon^* $, $ \lambda(\epsilon', \alpha_i) \leq \max_{\alpha}\lambda(\epsilon', \alpha) = \epsilon^*- \epsilon' <0, \forall i$, so from  \eqref{eq:aiApprox} and \eqref{eq:Rbounds}, $R \leq n (n+1)^2 \hspace{1mm}  2^{n(\epsilon^*- \epsilon')} \doteq 2^{n(\epsilon^*- \epsilon')} $ goes to $0$ exponentially fast.

 	Otherwise if $\epsilon' < \epsilon^* $, by the continuity of $\lambda(\epsilon', \alpha)$ in $\alpha$, for sufficiently large $n$, there is $\alpha_i$ for some $0\leq i \leq n-n_{lub}-1$ such that $\lambda(\epsilon', \alpha_i) \geq \frac{\max_{\alpha}\lambda(\epsilon', \alpha)}{2} >0$. Then
 	$R \geq \frac{1}{(n+1)^2} 2^{n \lambda(\epsilon', \alpha_i)} \doteq 2^{n \lambda(\epsilon', \alpha_i)}$ goes to infinity exponentially fast.

 \endproof

\textit{Proof of Corollary \ref{Coro:ratelosAndEpsi}:}

From \eqref{eq:epsilon'def}, 
\begin{equation*}
    \epsilon'= (1+\epsilon ) ( \frac{1-\delta}{2}  +o(1))-\frac{1}{2}= \frac{\epsilon}{2}-\frac{\delta}{2}(1+\epsilon)+o(1).
\end{equation*} 
Since $\delta>0$ can be chosen arbitrarily small, the conditions in theorem \ref{Thm:ratelossAndEpsi'} can be expressed in terms of $\epsilon$ as follows:
\begin{align*}
    R\to 0 \mbox{ exponentially fast when }  \epsilon' >\epsilon^* \iff \epsilon > 2\epsilon^* \\
    R\to \infty \mbox{ exponentially fast when }  \epsilon' < \epsilon^* \iff \epsilon < 2\epsilon^*
\end{align*}
\endproof

\subsection{Proof of Corollary \ref{Coro:LDGMwithUniformBound}:}
Since the code based on $G'$ uses a submatrix of $G'$ as its generator matrix, the column weights of this submatrix is upper bounded by $w_{u.b.}$ as well. From proposition \ref{prop:codeWithG'}, the probability of error of this code is upper bounded by that of the code based on $G$.
From lemma \ref{Lemma:newG_PeBound}, the code based on $G$ is capacity achieving, and, from corollary \ref{Coro:ratelosAndEpsi}, the rate loss $1+R$ goes to $1$ as $n$ grows large.

\subsection{Proofs for Subsection \ref{subsection:MainL>2}}

\textit{Proof of Lemma \ref{lemma:SparsityOrderGeneralL}:}

Note that 
\begin{align*}
    w_{MC}(n,G_l) &=  [(w_1\times w_2 \times \ldots \times w_l)^\frac{1}{l}]^n \\
    &= GM^n = (l^n)^{\log_l{GM} } = N^{\log_l{GM} },
\end{align*}
where $GM= GM(w_1, w_2 , \ldots , w_l)$ . Using \eqref{eq:logN'formula},  $[\log{N'}]^{\frac{1}{E(G_l)}} \leq   N \leq [\log{N'}]^{\frac{1}{{(1-\delta)E(G_l)}}}$ for sufficiently large $n$. 
Therefore, 
\begin{align*}
&[\log{N'}]^{  \frac{\log_l{GM}}{E(G_l)} }   \leq
w_{MC}(n,G_l) \leq 
[\log{N'}]^{\frac{\log_l{GM}}{{(1-\delta)E(G_l)}}  }.
\end{align*}
The sparsity order of the most common column weight can be bounded by:
\begin{equation}\label{eq:lambdaMCbound}
 \frac{\log_l{GM}} {E(G_l)}  \leq 
\lambda_{MC}(n,G_l) \leq 
\frac{\log_l{GM}}{{(1-\delta)E(G_l)}}.
\end{equation}
Writing  $GM$ and  $E(G_l)$ in terms of   $w_i$'s and $D_i$'s, we have
\[
 \frac{\log_l{GM}} {E(G_l)} 
 = \frac{ \sum_{i=1}^l \log_l w_i } {\sum_{i=1}^l \log_l D_i }. \label{eq:orderMC_bd}
\]
So equation \eqref{eq:lambdaMCbound} can be written as:
 \[
  \frac{ \sum_{i=1}^l \log_l w_i } {\sum_{i=1}^l \log_l D_i } \leq
\lambda_{MC}(n,G_l) \leq 
\frac{1}{1-\delta}
 \frac{ \sum_{i=1}^l \log_l w_i } {\sum_{i=1}^l \log_l D_i }.
\]
\endproof

\textit{Proof of Theorem  \ref{Thm:lambdaMC_lgeneral}:}

Let 
\[
G_l=
\left[
\begin{array}{c|c}
1 & \textbf{0}_{1,l-1} \\
\hline
\textbf{1}_{l-1,1} & I_{l-1}
\end{array}
\right]
\] be an $l\times l$ matrix. 

The geometric mean of column weights, $GM(w_1, \ldots, w_l )$, is $ GM(l, 1, \ldots, 1)= l^\frac{1}{l}$. The partial distances $\{D_i\}_{i=1}^l$ of $G_l$ are
$$
D_i=
\begin{cases}
2, \mbox{ for } i\geq 2\\
1, \mbox{ for } i =1.
\end{cases}
$$

It is simple to show that $ \lim_{ l \to \infty }\frac{\log_l{GM}} {E(G_l)} = 0 $. Hence for any fixed $\delta $, there is some $l^*$ such that $\frac{1}{(1-\delta)}\frac{\log_{l^*}{GM}}{{E(G_{l^*})}} <r $. 
By lemma \ref{lemma:SparsityOrderGeneralL}, $\lambda_{MC}(n,G_l) < r$ for sufficiently large $n$.
\endproof

\subsection{Proofs for Subsection \ref{sec:BSCmodel}}

\textit{Proof of Theorem \ref{Thm:queryBSC}:}

As in \cite{pang2019coding}, there is an $m\times n$ parity check matrix $H_n$, where $m= n[H_b(p)+\zeta(1-H_b(p))]$, for an LDPC code with row weight uniformly bounded by $O(\log({\frac{1}{\zeta}}))$\cite{sason2003parity,richardson2001capacity}.

Let the label vector be denoted by $\textbf{X} = (X_1, X_2, \ldots, X_n)^T \in \Ftwo^n$ and $G^*= G^*(n, \zeta, p)$ be the $K'\times N'$ submatrix of the $N'\times N'$ matrix $G'$, where $m= K' < N'C$, where $C$ is the channel capacity of a \BSC(q) channel, as described in corollary \ref{Coro:LDGMwithUniformBound}. 

Encode $H_n\textbf{X}$ with  $G^*$ to form  $\textbf{Y} = (H_n\textbf{X})^T G^* = \textbf{X}^T H_n^T G^*$. 
Since the code based on $G'$ is capacity achieving, the number of coded bits, i.e., the length of $\textbf{Y}$, approaches $\frac{m}{C} =  \frac{m}{1-H_b(q)}$ as $n$ grows large.
Each column of $G^*$ has weight upper bounded by
${[\log{N'}]}^{1+\epsilon}= {[O(\log m)]}^{1+\epsilon} ={[O(\log n)]}^{1+\epsilon}. $

Using a two-step encoding procedure as described above, we find upper bounds on the number of labels involved in the computation of a codebit and on the codeword length.
The first value can be bounded by \[O(\log({\frac{1}{\zeta}})) \times {O(\log n)}^{1+\epsilon} 
=O(
\log({\frac{1}{\zeta}})  {(\log n)}^{1+\epsilon}  ).\]
The length of $\textbf{Y}$, i.e., the number of rows of $ H_n^T G^*$, can be upper bounded by \[
(1+o(1))\frac{m}{1-H_b(q)} =(1+o(1))\frac{   n[H_b(p)+\zeta(1-H_b(p))]}{1-H_b(q)} \deff \, m'
\] for large $n$.

Assume $\mathbf{Z}$ is the received random vector when $\textbf{Y}$ is transmitted through a memoryless $\BSC(q)$.
Since the error probability of the code based on $G'$ vanishes as $n$ grows large, $H_n\textbf{X}$ can be recovered from $\mathbf{Z}$ with high probability.
From $H_n\textbf{X}$ $\textbf{X}$ can be decompressed with high probability, as discussed in \cite{pang2019coding}. 
To sum up, \textbf{X} can be decoded when $\textbf{Y}$ is transmitted through a memoryless $\BSC(q)$ with high probability.

Consider a query scheme consisting of $m'$ queries, each of which corresponds to a row of $ H_n^T G^*$. That is, the $i$th query includes label $X_j$ if and only if $ H_n^T G^*$ has a $1$ at position $(i,j)$, and the correct response to the $i$th query is $Y_i$.
For each query, at most $O(  \log \frac{1}{\zeta} {[\log n]}^{1+2\epsilon^*})$ items are sent to the crowdworker.
Since each query is assumed to be responded accurately with probability $1-q$, independent of others, the responses can be collected as a length-$m'$ vector and treated as $\mathbf{Z}$. From the discussion above, the label vector can be recovered from $\mathbf{Z}$ with high probability.
\endproof

\bibliographystyle{IEEEtran}
\bibliography{IEEEabrv}

\begin{thebibliography}{10}
\providecommand{\url}[1]{#1}
\csname url@samestyle\endcsname
\providecommand{\newblock}{\relax}
\providecommand{\bibinfo}[2]{#2}
\providecommand{\BIBentrySTDinterwordspacing}{\spaceskip=0pt\relax}
\providecommand{\BIBentryALTinterwordstretchfactor}{4}
\providecommand{\BIBentryALTinterwordspacing}{\spaceskip=\fontdimen2\font plus
\BIBentryALTinterwordstretchfactor\fontdimen3\font minus
  \fontdimen4\font\relax}
\providecommand{\BIBforeignlanguage}[2]{{%
\expandafter\ifx\csname l@#1\endcsname\relax
\typeout{** WARNING: IEEEtran.bst: No hyphenation pattern has been}%
\typeout{** loaded for the language `#1'. Using the pattern for}%
\typeout{** the default language instead.}%
\else
\language=\csname l@#1\endcsname
\fi
#2}}
\providecommand{\BIBdecl}{\relax}
\BIBdecl

\bibitem{gallager1962low}
R.~Gallager, ``Low-density parity-check codes,'' \emph{IRE Transactions on
  information theory}, vol.~8, no.~1, pp. 21--28, 1962.

\bibitem{arikan2009channel}
E.~Arikan, ``Channel polarization: A method for constructing capacity-achieving
  codes for symmetric binary-input memoryless channels,'' \emph{IEEE Trans.
  Inf. Theory}, vol.~55, no.~7, pp. 3051--3073, 2009.

\bibitem{karger2011iterative}
D.~R. Karger, S.~Oh, and D.~Shah, ``Iterative learning for reliable
  crowdsourcing systems,'' in \emph{Advances in neural information processing
  systems}, 2011, pp. 1953--1961.

\bibitem{vempaty2014reliable}
A.~Vempaty, L.~R. Varshney, and P.~K. Varshney, ``Reliable crowdsourcing for
  multi-class labeling using coding theory,'' \emph{IEEE Journal of Selected
  Topics in Signal Processing}, vol.~8, no.~4, pp. 667--679, 2014.

\bibitem{dimakis2010network}
A.~G. Dimakis, P.~B. Godfrey, Y.~Wu, M.~J. Wainwright, and K.~Ramchandran,
  ``Network coding for distributed storage systems,'' \emph{IEEE transactions
  on information theory}, vol.~56, no.~9, pp. 4539--4551, 2010.

\bibitem{lee2018speeding}
K.~Lee, M.~Lam, R.~Pedarsani, D.~Papailiopoulos, and K.~Ramchandran, ``Speeding
  up distributed machine learning using codes,'' \emph{IEEE Transactions on
  Information Theory}, vol.~64, no.~3, pp. 1514--1529, 2018.

\bibitem{mazumdar2017semisupervised}
A.~Mazumdar and S.~Pal, ``Semisupervised clustering, and-queries and locally
  encodable source coding,'' in \emph{Advances in Neural Information Processing
  Systems}, 2017, pp. 6489--6499.

\bibitem{pang2019coding}
C.-J. Pang, H.~Mahdavifar, and S.~S. Pradhan, ``Coding for crowdsourced
  classification with xor queries,'' \emph{Proceedings of IEEE Information
  Theory Workshop (ITW)}, 2019.

\bibitem{mackay1999good}
D.~J. MacKay, ``Good error-correcting codes based on very sparse matrices,''
  \emph{IEEE transactions on Information Theory}, vol.~45, no.~2, pp. 399--431,
  1999.

\bibitem{mackay1995good}
D.~J. MacKay and R.~M. Neal, ``Good codes based on very sparse matrices,'' in
  \emph{IMA International Conference on Cryptography and Coding}.\hskip 1em
  plus 0.5em minus 0.4em\relax Springer, 1995, pp. 100--111.

\bibitem{zhong2005approaching}
W.~Zhong, H.~Chai, and J.~Garcia-Frias, ``Approaching the shannon limit through
  parallel concatenation of regular {LDGM} codes,'' in \emph{Proceedings.
  International Symposium on Information Theory, 2005. ISIT 2005.}\hskip 1em
  plus 0.5em minus 0.4em\relax IEEE, 2005, pp. 1753--1757.

\bibitem{garcia2003approaching}
J.~Garcia-Frias and W.~Zhong, ``Approaching shannon performance by iterative
  decoding of linear codes with low-density generator matrix,'' \emph{IEEE
  Communications Letters}, vol.~7, no.~6, pp. 266--268, 2003.

\bibitem{zhong2005ldgm}
W.~Zhong and J.~Garcia-Frias, ``{LDGM} codes for channel coding and joint
  source-channel coding of correlated sources,'' \emph{EURASIP Journal on
  Applied Signal Processing}, vol. 2005, pp. 942--953, 2005.

\bibitem{LDGM_capAchieving2011}
A.~M. Kakhaki, H.~K. Abadi, P.~Pad, H.~Saeedi, K.~Alishahi, and F.~Marvasti,
  ``Capacity achieving random sparse linear codes,'' \emph{Preprint}, 2011.

\bibitem{mahdavifar2017scaling}
H.~Mahdavifar, ``Scaling exponent of sparse random linear codes over binary
  erasure channels,'' in \emph{2017 IEEE International Symposium on Information
  Theory (ISIT)}.\hskip 1em plus 0.5em minus 0.4em\relax IEEE, 2017, pp.
  689--693.

\bibitem{lin2018coding}
W.~Lin, S.~Cai, B.~Wei, and X.~Ma, ``Coding theorem for systematic {LDGM} codes
  under list decoding,'' in \emph{2018 IEEE Information Theory Workshop
  (ITW)}.\hskip 1em plus 0.5em minus 0.4em\relax IEEE, 2018, pp. 1--5.

\bibitem{korada2010polar}
S.~B. Korada, E.~Sasoglu, and R.~Urbanke, ``Polar codes: Characterization of
  exponent, bounds, and constructions,'' \emph{IEEE Transactions on Information
  Theory}, vol.~56, no.~12, pp. 6253--6264, 2010.

\bibitem{Arikan2}
E.~Ar{\i}kan, ``Source polarization,'' \emph{Proceedings of IEEE International
  Symposium on Information Theory (ISIT)}, pp. 899--903, 2010.

\bibitem{abbe2011polarization}
E.~Abbe, ``Polarization and randomness extraction,'' \emph{Proceedings of IEEE
  International Symposium on Information Theory (ISIT)}, pp. 184--188, 2011.

\bibitem{mondelli2015achieving}
M.~Mondelli, S.~H. Hassani, I.~Sason, and R.~L. Urbanke, ``Achieving {Marton's}
  region for broadcast channels using polar codes,'' \emph{IEEE Transactions on
  Information Theory}, vol.~61, no.~2, pp. 783--800, 2015.

\bibitem{goela2015polar}
N.~Goela, E.~Abbe, and M.~Gastpar, ``Polar codes for broadcast channels,''
  \emph{IEEE Transactions on Information Theory}, vol.~61, no.~2, pp. 758--782,
  2015.

\bibitem{STY}
E.~\c{S}a\c{s}o\u{g}lu, E.~Telatar, and E.~Yeh, ``Polar codes for the two-user
  binary-input multiple-access channel,'' \emph{IEEE Transactions on
  Information Theory}, vol.~59, no.~10, pp. 6583--6592, 2013.

\bibitem{MELK}
H.~Mahdavifar, M.~El-Khamy, J.~Lee, and I.~Kang, ``Achieving the uniform rate
  region of general multiple access channels by polar coding,'' \emph{IEEE
  Transactions on Communications}, vol.~64, no.~2, pp. 467--478, 2016.

\bibitem{MV}
H.~Mahdavifar and A.~Vardy, ``Achieving the secrecy capacity of wiretap
  channels using polar codes,'' \emph{IEEE Transactions on Information Theory},
  vol.~57, no.~10, pp. 6428--6443, 2011.

\bibitem{andersson2010nested}
M.~Andersson, V.~Rathi, R.~Thobaben, J.~Kliewer, and M.~Skoglund, ``Nested
  polar codes for wiretap and relay channels,'' \emph{IEEE Communications
  Letters}, vol.~14, no.~8, pp. 752--754, 2010.

\bibitem{mahdavifar2015polar}
H.~Mahdavifar, M.~El-Khamy, J.~Lee, and I.~Kang, ``Polar coding for
  bit-interleaved coded modulation,'' \emph{IEEE Transactions on Vehicular
  Technology}, vol.~65, no.~5, pp. 3115--3127, 2015.

\bibitem{karger2014budget}
D.~R. Karger, S.~Oh, and D.~Shah, ``Budget-optimal task allocation for reliable
  crowdsourcing systems,'' \emph{Operations Research}, vol.~62, no.~1, pp.
  1--24, 2014.

\bibitem{moser2019information}
S.~M. Moser, ``Information theory: Lecture notes,'' 2019.

\bibitem{cover2012elements}
T.~M. Cover and J.~A. Thomas, \emph{Elements of information theory}.\hskip 1em
  plus 0.5em minus 0.4em\relax John Wiley \& Sons, 2012.

\bibitem{sason2003parity}
I.~Sason and R.~Urbanke, ``Parity-check density versus performance of binary
  linear block codes over memoryless symmetric channels,'' \emph{IEEE
  Transactions on Information Theory}, vol.~49, no.~7, pp. 1611--1635, 2003.

\bibitem{richardson2001capacity}
T.~J. Richardson and R.~L. Urbanke, ``The capacity of low-density parity-check
  codes under message-passing decoding,'' \emph{IEEE Transactions on
  information theory}, vol.~47, no.~2, pp. 599--618, 2001.

\end{thebibliography}

\end{document}